\documentclass[11pt]{article}%\pdfoutput=1
\usepackage{cite}
\usepackage{scalerel}
\usepackage{xcolor}
\usepackage{shuffle}
\usepackage{amsmath,amsfonts,amssymb}
\usepackage{latexsym,epsfig}
\usepackage{dsfont}
\usepackage{hyperref}
\usepackage{extarrows}
\usepackage{comment} % To comment several lines at once.
\usepackage{tikz-cd}

\def\hybrid{
        \topmargin -20pt
        \oddsidemargin 0pt
        \headheight 0pt \headsep 0pt
        \textwidth 6.25in % A4 paper
        \textheight 9.5in % A4 paper
        \marginparwidth .875in
        \parskip 5pt plus 1pt \jot = 1.5ex}

\hybrid

 \csname
@addtoreset\endcsname{equation}{section}

\def\cC{{\cal C}}
\def\cB{{\cal B}}
\def\cG{{\cal G}}
\def\cJ{{\cal J}}

\def\cD{\mathcal{D}}
\def\cF{{\cal F}}
\def\cO{{\cal O}}
\def\cA{{\cal A}}
\def\cE{{\cal E}}
\def\cM{{\cal M}}
\def\cN{{\cal N}}

\def\cV{{\cal V}}

\def\cX{{\cal X}}
\def\cH{{\cal H}}

\def\del{\partial}
\def\l{\langle}
\def\r{\rangle}

\def\B{\square}

\allowdisplaybreaks

\def\bpm{\begin{pmatrix}}
\def\epm{\end{pmatrix}}
\newcommand{\bra}[1]{\langle#1\rvert}
\newcommand{\ket}[1]{\lvert#1\rangle}

\begin{document}

\begin{titlepage}

\rightline{June 2024}
\rightline{HU-EP-24/20}
  
\begin{center}
\vskip 1.5cm
{\Large \bf{Yang-Mills theory from the worldline}}
\vskip 1.7cm

{\large\bf {Roberto Bonezzi}}
\vskip 1.6cm
{\it  Institute for Physics, Humboldt University Berlin,\\
 Zum Gro\ss en Windkanal 6, D-12489 Berlin, Germany}\\[1.5ex]  
roberto.bonezzi@physik.hu-berlin.de

\vskip .1cm

\vskip .2cm

\end{center}

\bigskip\bigskip
\begin{center} 
\textbf{Abstract}

\end{center} 
\begin{quote}

We construct off-shell vertex operators for the bosonic spinning particle. Using the language of homotopy algebras, we show that the full nonlinear structure of Yang-Mills theory, including its gauge transformations, is encoded in the commutator algebra of the worldline vertex operators. To do so, we deform the worldline BRST operator by coupling it to a background gauge field and show that the coupling is consistent on a suitable truncation of the Hilbert space. On this subspace, the square of the BRST operator is proportional to the Yang-Mills field equations, which we interpret as an operator Maurer-Cartan equation for the background. This allows us to define further vertex operators in different ghost numbers, which correspond to the entire $L_\infty$ algebra of Yang-Mills theory.
Besides providing a precise map of a fully nonlinear field theory into a worldline model, we expect these results will be valuable to investigate the kinematic algebra of Yang-Mills, which is central to the double copy program.

\end{quote} 
\vfill
\setcounter{footnote}{0}
\end{titlepage}

\tableofcontents
%\newpage

\vspace{5mm}

\section{Introduction}

The first-quantized description of quantum field theory observables, such as the worldsheet computation of string amplitudes, often reveals structures that are hidden in a more conventional field theoretic approach. Paramount among these is the double copy construction of gravity amplitudes in terms of gauge theory amplitudes. First discovered in the context of first-quantized string theory \cite{Kawai:1985xq}, starting from the seminal papers \cite{Bern:2008qj,Bern:2010ue} by Bern, Carrasco and Johansson it has flourished in a number of directions in quantum field theory and is now a prominent aspect of the modern amplitude program (see e.g. the reviews \cite{Bern:2019prr,Bern:2022wqg,Adamo:2022dcm}). 

Similar to the worldsheet approach to string theory, the worldline formalism is a first-quantized description of relativistic point particles.
Pioneered by Feynman \cite{Feynman:1950ir,Feynman:1951gn}, it gained attention with the introduction of the Bern-Kosower \cite{Bern:1990cu,Bern:1992cz} rules.  These were originally derived from the point particle limit of string theory and provided compact master formulas for $n$-gluon one-loop amplitudes in QCD. Soon after \cite{Strassler:1992zr}, Strassler showed that the Bern-Kosower rules follow from a genuine worldline path integral on the circle. Since then, the worldline formalism has been extended to describe various couplings \cite{DHoker:1995uyv,Schubert:2001he}, including scalars \cite{Bastianelli:2002fv}, spinors \cite{Bastianelli:2002qw} and $p$-forms \cite{Bastianelli:2005vk} coupled to gravity.
More recently, worldline techniques have been applied to address the double copy \cite{Ahmadiniaz:2021fey,Ahmadiniaz:2021ayd,Shi:2021qsb} and
have found new applications in the context of gravitational wave physics \cite{Mogull:2020sak,Jakobsen:2021zvh}. 

In this first-quantized approach, spacetime spin is generated by adding internal degrees of freedom to the particle, in terms of either worldline fermions or bosons \cite{Barducci:1976xq,Berezin:1976eg,Gershun:1979fb,Howe:1988ft,Hallowell:2007qk,Bastianelli:2009eh}.
In order to preserve unitarity, these extra degrees of freedom come together with local (super)symmetries on the worldline. Upon canonical quantization, one can encode free field equations and their gauge symmetry in target space via the worldline BRST system \cite{Henneaux:1987cp,Barnich:2003wj,Barnich:2004cr,Cherney:2009mf}
\begin{equation}
Q\ket{\psi}=0\;,\quad\delta\ket{\psi}=Q\ket{\Lambda}\;,   
\end{equation}
where $Q$ is the first quantized BRST charge and spacetime fields are contained in the BRST wave function $\ket{\psi}$.
The free graviton, for instance, can be described by a worldline with $\cN=4$ supersymmetry. Interestingly, its internal degrees of freedom are given by two copies of the ones of the $\cN=2$ particle, describing a free gluon. In this respect, their BRST quantization  naturally leads to an off-shell and gauge invariant double copy for the free target space theories, relating the Maxwell and Fierz-Pauli lagrangians. This feature inspired a double copy program, based on the framework of homotopy algebras, that has led to a gauge invariant and off-shell double copy\footnote{For other approaches to off-shell double copy constructions see, e.g. \cite{Bern:2010yg,Anastasiou:2018rdx,Borsten:2020xbt,Borsten:2020zgj,Ferrero:2020vww,Borsten:2021hua,Diaz-Jaramillo:2021wtl,Godazgar:2022gfw}.} of Yang-Mills theory up to quartic order \cite{Bonezzi:2022yuh,Bonezzi:2022bse}.

The spinning particles naturally describe free gauge theories in spacetime, but there is no systematic procedure to construct nonlinear theories. While it is well known that interactions of scalars and spinors with gauge fields and gravitons are represented by inserting vertex operators on the worldline, the self-interactions of pure Yang-Mills and gravity are much less understood. The first important progress in this direction was made in \cite{Dai:2008bh}, where the authors found a consistent coupling of the $\cN=2$ particle to a background Yang-Mills field. Similarly, the coupling of the $\cN=4$ particle to background gravity was achieved in \cite{Bonezzi:2018box}, where it was shown that consistency of the worldline quantum theory demands that the background obeys Einstein's equations. This led to identify the correct path integral of the $\cN=4$ particle on the circle in \cite{Bastianelli:2019xhi}, which was used in \cite{Bastianelli:2022pqq,Bastianelli:2023oca} to reproduce the one-loop divergences of Einstein gravity in four and six dimensions. The setup of \cite{Bonezzi:2018box} was generalized in \cite{Bonezzi:2020jjq} to include couplings to the Kalb-Ramond two-form and dilaton. Constraints on consistent backgrounds from nilpotence of the BRST operator were also studied in similar contexts in \cite{Grigoriev:2021bes,Carosi:2021wbi}.

Despite the progress in coupling the gluon and graviton to their respective backgrounds, the precise relation between the worldline description and the nonlinear field theories remains an open problem. 
In this paper we bridge the gap for the case of Yang-Mills theory. Specifically, we will map the full nonlinear structure of Yang-Mills, including its gauge transformations, to the algebra of off-shell vertex operators acting on the BRST Hilbert space of a spinning particle. 

To this end, we first couple the bosonic spinning particle \cite{Cherney:2009mf,Bastianelli:2009eh} to a Yang-Mills background by deforming its BRST operator. The spacetime spectrum of the worldline includes massless particles of all integer spins. We show that the coupling is consistent on the spin one sector of the Hilbert space, where the
deformed BRST operator $Q_A$ is nilpotent if the background satisfies the nonlinear Yang-Mills equations. Rather than interpreting this as a condition on possible backgrounds, we think of $Q_A^2=0$ itself as an operator equation of motion for the gauge field in $Q_A$. This allows us to determine off-shell vertex operators, starting from the expansion
\begin{equation}
Q_A=Q+\cV(A)+\tfrac12\,\cV_2(A,A) \;,   
\end{equation}
not only for the fields, but for gauge parameters and equations of motion as well. Throughout this analysis we will use the language of homotopy Lie (or $L_\infty$) algebras \cite{Zwiebach:1992ie,Lada:1992wc,Hohm:2017pnh,Jurco:2018sby}, as it streamlines the nonlinear structures of gauge theories in terms of relations between multilinear brackets. We will give the necessary background material in the body of the paper. For the specific case of Yang-Mills theory, the $L_\infty$ structure was first identified in \cite{Zeitlin:2007vv} from open string vertex operators \cite{Zeitlin:2007vd}. 

Having established a precise map between the Yang-Mills $L_\infty$ brackets and commutators of vertex operators\footnote{In \cite{Zeitlin:2008cc} the $L_\infty$ brackets where used to \emph{define} a deformed differential.}, we are able to clarify the role of the nonlinear terms appearing in $Q_A$. In particular, we show how the bilinear vertex operator $\cV_2(A,A)$ determines the quartic coupling of the theory, from which we recover the full Yang-Mills action as a worldline expectation value: 
\begin{equation}
S_{\rm YM}[A]=\frac12\,\big\l\cV(A)Q\cV(A)\big\r+\frac13\,\big\l\cV^3(A)\big\r+\frac18\,\big\l\cV(A)\{\cV_2(A,A),\cV(A)\}\big\r \;.   
\end{equation} 
The dictionary established in this paper, relating the $L_\infty$ algebra of Yang-Mills to the algebra of vertex operators, should serve as a valuable starting point for the investigation of the off-shell kinematic algebra identified in \cite{Reiterer:2019dys,Ben-Shahar:2021zww}, which is central to the algebraic double copy program pursued in \cite{Bonezzi:2022bse,Borsten:2022vtg,Borsten:2023reb,Bonezzi:2023pox,Borsten:2023ned,Borsten:2023paw,Bonezzi:2023lkx,Bonezzi:2024dlv}.

The rest of this paper is organized as follows. In section \ref{sec:spinning particle} we review the bosonic spinning particle and its BRST quantization, emphasizing the target space interpretation in terms of the $L_\infty$ algebra of a free gauge theory. In section \ref{sec:background coupling} we introduce the coupling to a background gauge field by deforming the BRST operator. We show it is nilpotent, when the background is on-shell, upon restricting to the spin one sector of the Hilbert space. We use this in section \ref{sec:vertex operators} to interpret $Q_A^2$ as an operator Maurer-Cartan equation, from which we identify the off-shell vertex operators. Comparing the operator algebra with the $L_\infty$ relations of Yang-Mills, we fix the dictionary between the two and derive the action as a first-quantized expectation value. We close in section \ref{sec:conclusions} with a brief outlook on future directions.

\section{The bosonic spinning particle and free massless fields}\label{sec:spinning particle}

In this section we will review how the quantization of the bosonic spinning worldline gives rise to massless particles of arbitrary spin in spacetime \cite{Hallowell:2007qk,Bastianelli:2009eh,Cherney:2009mf}. We will emphasize the BRST quantization of the theory and its target space interpretation. In particular, at the end of the section we will relate the worldline BRST quantization to the $L_\infty$ description of free gauge field theories in spacetime.

In order to construct the worldline action, we start from the following symplectic term:
\begin{equation}
S_{\rm symp}=\int d\tau\,\Big[p_\mu\dot x^\mu-i\,\bar\alpha^\mu\dot\alpha_\mu\Big]\;, 
\end{equation}
where $\mu=0,\ldots,D-1$ is a target space Lorentz index and $\bar\alpha^\mu=(\alpha^\mu)^*$. The phase space thus consists of the standard coordinates and momenta $(x^\mu,p_\nu)$, augmented by the complex bosonic pair $(\alpha^\mu,\bar\alpha^\nu)$. The latter can be thought of as a worldline analog of open string modes $\alpha^\mu_{\pm1}$.  
We now introduce the following triplet of phase space functions:
\begin{equation}
H:=\frac12\,p^2\;,\quad L:=\alpha^\mu p_\mu \;,\quad \bar L:=\bar\alpha^\mu p_\mu\;,  
\end{equation}
which form a closed algebra under Poisson brackets.
$H$ is the Hamiltonian for $\tau$ translations, while $L$ and $\bar L$ mix $x^\mu$ with $\alpha^\mu$ and $\bar \alpha^\mu$, respectively. The functions $H$, $L$ and $\bar L$ are analogous to the $L_0$ and $L_{\pm1}$ Virasoro modes of the bosonic open string. In fact, they can be obtained from a contraction of the $sl(2,\mathbb{R})$ subalgebra of Virasoro in the tensionless limit $\alpha'\rightarrow\infty$ \cite{Bengtsson:1986ys,Bouatta:2004kk}.

We will interpret the states of the quantum theory as spacetime massless particles, with spin degrees of freedom associated to the oscillators $\alpha^\mu$. To this end, one needs to gauge the Hamiltonian $H$ to enforce the mass-shell condition, as well as the ``Virasoro charges'' $L$ and $\bar L$. Gauging the latter is necessary in order to remove unphysical degrees of freedom associated to oscillators $\alpha^\pm$ in lightcone directions. The worldline model is thus described by the action
\begin{equation}
S=\int d\tau\,\Big[p_\mu\dot x^\mu-i\,\bar\alpha^\mu\dot\alpha_\mu-e\,H-\bar u\,L-u\,\bar L\Big] \;,
\end{equation}
which is invariant under $\tau$ reparametrizations and local ``Virasoro transformations'' generated by $L$ and $\bar L$:
\begin{equation}
\begin{array}{ll}
\delta x^\mu=\epsilon\,p^\mu+\xi\,\bar\alpha^\mu+\bar\xi\,\alpha^\mu\;,&  \delta p_\mu=0\;,\\
\delta\alpha^\mu=i\,\xi\,p^\mu\;,&\delta\bar\alpha^\mu=-i\,\bar\xi\,p^\mu\;,\\
\;\;\delta u=\dot\xi\;,\qquad\delta\bar u=\dot{\bar\xi}\;,&\;\;\delta e=\dot\epsilon+2i\,u\,\bar\xi-2i\,\bar u\,\xi\;,
\end{array}    
\end{equation}
with local parameters $\epsilon(\tau)$ and $\xi(\tau)$, with $\bar\xi=\xi^*$.
The Lagrange multipliers $e(\tau)$ and complex $u(\tau)$ and $\bar u(\tau)$ can be viewed as a triplet of einbeins and enforce the classical constraints $H=L=\bar L=0$.
We now turn to the quantum mechanical treatment of this constrained system, starting from Dirac quantization.

\subsection{Dirac quantization: gauge fixed spacetime theory}

Upon canonical quantization, the symplectic structure gives rise to the following commutation relations:
\begin{equation}
[x^\mu,p_\nu]=i\,\delta^\mu_\nu \;,\quad [\bar\alpha^\mu,\alpha^\nu]=\eta^{\mu\nu}\;,   
\end{equation}
yielding the quantum constraint algebra
\begin{equation}\label{quantum constr algebra}
[\bar L,L]=2\,H\;,\quad [H,L]=0\;,\quad [H,\bar L]=0\;,  
\end{equation}
where for operators we use the same symbols as for their classical counterparts: $H=\frac12\,p^2$, $L=\alpha^\mu p_\mu$, $\bar L=\bar\alpha^\mu p_\mu$.
As Hilbert space we choose the tensor product of smooth functions of $x^\mu$ with power series in $\alpha^\mu$.
The latter can be viewed as the Fock space constructed with creation operators $\alpha^\mu$ on a vacuum state $\ket{0}$ annihilated by $\bar\alpha^\mu$.
A generic state thus takes the form
\begin{equation}
\ket{\varphi}=\sum_{s=0}^\infty\ket{\varphi_s}\;,\quad\ket{\varphi_s}=\frac{1}{s!}\,\varphi_{\mu_1\ldots\mu_s}(x)\,\alpha^{\mu_1}\cdots\alpha^{\mu_s}\ket{0}\;,  
\end{equation}
which is interpreted as a collection of spacetime symmetric tensor fields of arbitrary rank $s$.
On this space $p_\mu$ and $\bar\alpha^\mu$ act as derivative operators:
\begin{equation}
p_\mu=-i\,\del_\mu\;,\quad\bar\alpha^\mu=\eta^{\mu\nu}\frac{\del}{\del\alpha^\nu}\;,    
\end{equation}
upon identifying the ket $\alpha^{\mu_1}\cdots\alpha^{\mu_s}\ket{0}$ with the monomial $\alpha^{\mu_1}\cdots\alpha^{\mu_s}$.
This yields the following representation for the quantum constraints:
\begin{equation}\label{constraints}
H=-\frac{1}{2}\,\B \;,\quad L=-i\alpha^\mu\del_\mu\;,\quad\bar L=-i\frac{\del}{\del\alpha^\mu}\del^\mu \;,  
\end{equation}
where $\B=\del^\mu\del_\mu$ is the wave operator. $L$ and $\bar L$ act on symmetric tensors as the symmetrized gradient and divergence, respectively:
\begin{equation}
iL\ket{\varphi_s}= \frac{1}{s!}\,\del_{(\mu_1}\varphi_{\mu_2\ldots\mu_{s+1})}\,\alpha^{\mu_1}\cdots\alpha^{\mu_{s+1}}\ket{0}\;,\quad i\bar L\ket{\varphi_s}= \frac{1}{(s-1)!}\,\del^\nu\varphi_{\nu\mu_2\ldots\mu_{s}}\,\alpha^{\mu_2}\cdots\alpha^{\mu_{s}}\ket{0}   \;.
\end{equation}

Declaring that $(\alpha^\mu)^\dagger=\bar\alpha^\mu$ allows us to define a bra state, and thus an inner product, as
\begin{equation}\label{Dirac inner}
\begin{split}
\bra{\varphi_s}&=\frac{1}{s!}\,\varphi^*_{\mu_1\ldots\mu_s}(x)\,\bra{0}\bar\alpha^{\mu_1}\cdots\bar\alpha^{\mu_s} \;,\\
\l\chi_{s'}|\varphi_s\r&=\frac{1}{s!s'!}\int d^Dx\,\chi^*_{\mu_1\ldots\mu_{s'}}\varphi_{\nu_1\ldots\nu_s}\,\bra{0}\bar\alpha^{\mu_1}\cdots\bar\alpha^{\mu_{s'}}\,\alpha^{\nu_1}\cdots\alpha^{\nu_s}\ket{0}\\
&=\delta_{ss'}\,\frac{1}{s!}\int d^Dx\,\chi^*_{\mu_1\ldots\mu_s}\varphi^{\mu_1\ldots\mu_s}\;.
\end{split}   
\end{equation}
For the $x-$dependent part we chose the usual quantum mechanical inner product, ensuring that $p_\mu^\dagger=p_\mu$. This implies that $H$ is self-adjoint, while $L^\dagger=\bar L$. 

We now proceed with the Dirac quantization, in which the quantum constraints select a physical subspace of the Hilbert space, which we denote by $\cH_{\rm phys}$. This is determined by requiring that the constraints have vanishing matrix elements with physical states:
\begin{equation}\label{Dirac}
\bra{\chi}(H,L,\bar L)\ket{\psi}=0\quad\forall\;\chi,\psi\in\cH_{\rm phys}\;.    
\end{equation}
Given that $H$ is self-adjoint, while $L^\dagger=\bar L$,
we define the physical state condition by
\begin{equation}\label{phys state conditions}
\ket{\varphi}\in\cH_{\rm phys}\quad\longleftrightarrow \quad H\ket{\varphi}=0\;,\quad\bar L\ket{\varphi}=0\;,
\end{equation}
which is sufficient to ensure that \eqref{Dirac} holds for $L$ as well.
The physical state conditions \eqref{phys state conditions} govern the dynamics of the system completely: since the Hamiltonian is itself a constraint, the Schr\"odinger equation is trivially solved by demanding that physical states do not depend on the worldline parameter $\tau$.

In terms of spacetime fields of rank $s$ the condition \eqref{phys state conditions} amounts to
\begin{equation}\label{Dirac eom}
\B\varphi_{\mu_1\ldots\mu_s}=0\;,\quad\del^\nu\varphi_{\nu\mu_1\ldots\mu_{s-1}}=0\;,    
\end{equation}
meaning that physical states are massless and transverse. These conditions alone are not enough to remove all unphysical polarizations. To do so one has to take into account that the above equations are invariant under the on-shell gauge transformation
\begin{equation}\label{on-shell gauge}
\delta\varphi_{\mu_1\ldots\mu_s}=s\,\del_{(\mu_1}\xi_{\mu_2\ldots\mu_s)}\;,   
\end{equation}
with an on-shell and transverse gauge parameter: $\del^\nu\xi_{\nu\mu_2\ldots\mu_{s-1}}=0$, $\B\xi_{\mu_1\ldots\mu_{s-1}}=0$.
This is precisely enough to remove all unphysical components. Since the tensor field $\varphi_{\mu_1\ldots\mu_s}$ is \emph{not} traceless, the above equations propagate a reducible spectrum of massless particles\footnote{We remind the reader that the physical polarizations of a massless particle of spin $s$ form the rank $s$ symmetric traceless representation of the little group $SO(D-2)$.}. For fixed rank $s$, $\varphi_{\mu_1\ldots\mu_s}$ propagates massless spin $s$, $s-2$, $s-4$ etc, down to spin one or zero. The spectrum is irreducible for $s=1$, where the physical state conditions reduce to the Maxwell equations in Lorenz gauge:
\begin{equation}
\B A_\mu=0\;,\quad\del^\mu A_\mu=0\;,    
\end{equation}
together with the on-shell gauge symmetry $\delta A_\mu=\del_\mu\lambda$, with $\B\lambda=0$.

In terms of the Dirac constrained system \eqref{phys state conditions}, the on-shell gauge symmetry \eqref{on-shell gauge} is interpreted as the appearance of null states of the form
\begin{equation}
\ket{\varphi_{\rm null}}=L\ket{\xi}\;,\quad H\ket{\xi}=\bar L\ket{\xi}=0\;.    
\end{equation}
These states are physical, but have zero norm and zero overlap with any other physical state. The space of nontrivial physical states is thus the equivalence class $\ket{\varphi}\sim\ket{\varphi}+L\ket{\xi}$, which reproduces the on-shell gauge symmetry discussed above.
The free field theory described by \eqref{Dirac eom} and \eqref{on-shell gauge} is (partially) gauge fixed and non-Lagrangian. In the following we will obtain a gauge invariant and Lagrangian formulation from BRST quantization.

\subsection{BRST quantization: gauge invariant spacetime theory}

We will now treat the constraint algebra \eqref{quantum constr algebra} in the Hamiltonian BRST framework, where physical states are identified as elements of the BRST cohomology. In general, given a set $\{G_i\}$ of quantum Hamiltonian constraints forming a Lie algebra
\begin{equation}\label{first class Lie}
[G_i,G_j]=f_{ij}^k\,G_k\;,    
\end{equation}
one proceeds by assigning a ghost conjugate pair to each constraint:
\begin{equation}\label{ghosts generic}
G_i\;\rightarrow\;(b_i,c^i)\;,\quad\{b_i,c^j\}=\delta_i^j\;,    
\end{equation}
where the $c^i$ and $b_i$ have ghost number $+1$ and $-1$, respectively. One can then construct a ghost number one BRST operator via
\begin{equation}
Q:=c^i\,G_i-\frac12\,f_{ij}^k\,c^ic^j\,b_k\;,  
\end{equation}
which is nilpotent thanks to the commutation relations \eqref{first class Lie}, \eqref{ghosts generic} and Jacobi identity of the structure constants $f_{ij}^k$. On the larger BRST Hilbert space (given by tensoring the ``matter'' and ghost sectors), the BRST cohomology agrees with the Dirac quantization discussed in the previous section.

Applying this procedure to the constraint algebra \eqref{quantum constr algebra}, we introduce the ghost pairs:
\begin{equation}
\begin{split}
H\;&\rightarrow\;(b,c)\;,\quad\{b,c\}=1\;,\\
L\;&\rightarrow\;(\cB,\bar\cC)\;,\quad\{\cB,\bar\cC\}=1\;,\\  \bar L\;&\rightarrow\;(\bar\cB,\cC)\;,\quad\{\bar\cB,\cC\}=1\;,
\end{split}    
\end{equation}
where $(c,\cC,\bar\cC)$ have ghost number $+1$ and $(b,\cB,\bar\cB)$ have ghost number $-1$.
All ghosts are Grassmann odd and anticommutators not displayed above vanish.
The BRST operator is then given by
\begin{equation}
Q:=c\,\B+(\bar\cC\,\alpha^\mu+\cC\,\bar\alpha^\mu)\del_\mu-\cC\,\bar\cC\,b\;, \quad Q^2=0\;,  
\end{equation}
where we identified the momentum operator with the spacetime derivative $p_\mu\equiv-i\del_\mu$.

We now come to construct the BRST-extended Hilbert space $\cH$. This is the tensor product of the Hilbert space $\cH_{\rm matter}$ associated to the $(x^\mu,p_\mu,\alpha^\mu,\bar\alpha^\mu)$ operators with the ghost Hilbert space $\cH_{\rm gh}$. Since all ghosts are Grassmann odd, $\cH_{\rm gh}$ is finite dimensional.
We choose the ghost vacuum $\ket{0}_{\rm gh}$ to be annihilated by $b$, $\bar\cB$ and $\bar\cC$. The ghost Hilbert space is then given by acting (at most once) on this vacuum with the creation operators $c$, $\cC$ and $\cB$.
Altogether, denoting by $\ket{0}$ the full BRST vacuum we have
\begin{equation}\label{vacuum}
(\bar\alpha^\mu,b,\bar\cB,\bar\cC)\ket{0}=0\;.    
\end{equation}
A generic state in $\cH$ can thus be written as
\begin{equation}\label{Psi BRST}
\ket{\psi}=\sum_{s=0}^\infty\sum_{p,q,r=0}^1 c^p\,\cC^q\,\cB^r\ket{\psi_{s,p,q,r}}\;,\quad   \ket{\psi_{s,p,q,r}}=\frac{1}{s!}\,\psi_{\mu_1\ldots\mu_s}^{(p,q,r)}(x)\,\alpha^{\mu_1}\cdots\alpha^{\mu_s}\ket{0} \;,
\end{equation}
with the annihilation operators acting via derivatives
\begin{equation}
\bar\alpha_\mu=\frac{\del}{\del\alpha^\mu}\;,\quad b=\frac{\del}{\del c}\;,\quad\bar\cB=\frac{\del}{\del\cC}\;,\quad\bar\cC=\frac{\del}{\del\cB}\;,    
\end{equation}
on polynomials in $(\alpha^\mu,c, \cC,\cB)$.
The inner product \eqref{Dirac inner} is extended to the ghost sector of $\cH$ by the following hermiticity assignments:
\begin{equation}\label{ghostherm}
c^\dagger=c\;,\quad b^\dagger=b\;,\quad \cC^\dagger=-\bar\cC\;,\quad\cB^\dagger=-\bar\cB\;, 
\end{equation}
ensuring that $Q^\dagger=Q$.
Since $c$ and $b$ are self-adjoint, the overlap of the vacuum with itself vanishes\footnote{One has $\l0|0\r=\bra{0}cb+bc\ket{0}=0$ upon using $b^\dagger=b$ and $b\ket{0}=0$. This is typical of $bc$ systems arising from reparametrization invariance, with ghost zero modes associated to Killing vectors.} and we normalize the basic overlap to be
\begin{equation}\label{overlap}
\bra{0}c\ket{0}=1\;.    
\end{equation}

The Hilbert space $\cH$ can be decomposed according to two integer degrees. To this end, we define the ghost number operator $\cG$ and the $U(1)$ charge $\cJ$ via
\begin{equation}\label{G N}
\begin{split}
\cG&:=cb+\cC\bar\cB-\cB\bar\cC=N_c+N_\cC-N_\cB\;,\\
\cJ&:=\alpha^\mu\bar\alpha_\mu+\cC\bar\cB+\cB\bar\cC=N_\alpha+N_\cC+N_\cB\;,
\end{split}
\end{equation}
where the $N_i$ count the number of the corresponding oscillators, so that the charge $\cJ$ counts the total occupation number. $\cJ$ and $\cG$ can be diagonalized simultaneously, since $[\cJ,\cG]=0$, decomposing $\cH$ into the double direct sum
\begin{equation}
\cH=\bigoplus_{s=0}^\infty\bigoplus_{k=-1}^{2}\cH_{s,k}\;, 
\end{equation}
in terms of eigenstates with $\cJ=s$ and $\cG=k$. The BRST operator obeys
\begin{equation}
[\cG,Q]=Q\;,\quad [\cJ,Q]=0\;,    
\end{equation}
implying that it acts as a map $Q:\cH_{s,k}\rightarrow\cH_{s,k+1}$.
The BRST cohomology can thus be studied separately at any fixed value of $s$, which coincides with the maximal spin being propagated. This will be instrumental for coupling the theory to a Yang-Mills background in the next section. 

We now restrict to the subspace with $\cJ=s$ fixed but arbitrary and determine the BRST cohomology at ghost number zero. We thus consider the Hilbert subspace $\cH_s=\bigoplus_{k=-1}^2\cH_{s,k}$, where $k$ labels the ghost number. The ``string field'' at ghost number zero is given by
\begin{equation}
\begin{split}
\ket{\psi_s}&=\ket{\varphi_s}+c\cB\,\ket{f_{s-1}}+\cC\cB\,\ket{\chi_{s-2}}  \;,\qquad{\rm with}\\
\ket{\varphi_{s}}&=\frac{1}{s!}\,\varphi_{\mu_1\ldots\mu_s}(x)\,\alpha^{\mu_1}\cdots\alpha^{\mu_s}\ket{0}\;,\quad \ket{f_{s-1}}=\frac{1}{(s-1)!}\,f_{\mu_1\ldots\mu_{s-1}}(x)\,\alpha^{\mu_1}\cdots\alpha^{\mu_{s-1}}\ket{0}\;,\\
\ket{\chi_{s-2}}&=\frac{1}{(s-2)!}\,\chi_{\mu_1\ldots\mu_{s-2}}(x)\,\alpha^{\mu_1}\cdots\alpha^{\mu_{s-2}}\ket{0}\;,
\end{split}  
\end{equation}
where in the first line we displayed explicitly the ghost dependence. This triplet of fields is usually obtained in string-like formulations of higher spin fields \cite{Bouatta:2004kk}. Here $\varphi_s$ is the (reducible) spin $s$ field, $f_{s-1}$ is an auxiliary field and $\chi_{s-2}$ can be viewed as a spin $s-2$ dilaton.

The BRST closure condition $Q\ket{\psi_{s}}=0$ is interpreted as the field equations
\begin{equation}
\begin{split}
\B\varphi_{\mu_1\ldots\mu_s}-s\,\del_{(\mu_1}\,f_{\mu_2\ldots\mu_{s)}}&=0\;,\\
\B\chi_{\mu_1\ldots\mu_{s-2}}-\del^\rho f_{\rho\mu_1\ldots\mu_{s-2}}&=0\;,\\
\del^\rho\varphi_{\rho\mu_1\ldots\mu_{s-1}}-(s-1)\,\del_{(\mu_1}\chi_{\mu_2\ldots\mu_{s-1})}-f_{\mu_1\ldots\mu_{s-1}}&=0\;,     
\end{split}   
\end{equation}
which shows that the field $f_{s-1}$ is auxiliary.
Spacetime gauge symmetry is then viewed as the equivalence relation $\ket{\psi_{s}}\sim\ket{\psi_{s}}+Q\ket{\Lambda_{s}}$, where the gauge parameter $\ket{\Lambda_{s}}$ has ghost number $-1$ and $\cJ=s$:
\begin{equation}
\ket{\Lambda_{s}}=\cB\,\ket{\xi_{s-1}}\;,\quad\ket{\xi_{s-1}}=\frac{1}{(s-1)!}\,\xi_{\mu_1\ldots\mu_{s-1}}(x)\,\alpha^{\mu_1}\cdots\alpha^{\mu_{s-1}}\ket{0}  \;. 
\end{equation}
The resulting gauge transformations for the component fields are given by
\begin{equation}
\delta\varphi_{\mu_1\ldots\mu_s}=s\,\del_{(\mu_1}\xi_{\mu_2\ldots\mu_{s})}\;,\quad \delta\chi_{\mu_1\ldots\mu_{s-2}}=\del^\rho\xi_{\rho\mu_1\ldots\mu_{s-2}}\;,\quad\delta f_{\mu_1\ldots\mu_{s-1}}=\B\xi_{\mu_1\ldots\mu_{s-1}}\;.   
\end{equation}
One can make contact with Dirac quantization in a two-step gauge fixing: first one uses the off-shell gauge symmetry to fix $f_{\mu_1\ldots\mu_{s-1}}=0$. This leaves residual gauge transformations with a parameter obeying $\B\xi_{\mu_1\ldots\mu_{s-1}}=0$. One further uses the divergence of the residual parameter to fix $\chi_{\mu_1\ldots\mu_{s-2}}=0$ on-shell. At this point one is left with $\B\varphi_{\mu_1\ldots\mu_{s}}=0$, $\del^    \rho\varphi_{\rho\mu_1\ldots\mu_{s-1}}=0$ with residual harmonic and transverse gauge parameter, as in the Dirac procedure.

Using the inner product on $\cH$ one can derive the gauge invariant field equations $Q\ket{\psi_s}=0$ from the variation of a string field theory-like action \cite{Bouatta:2004kk}:
\begin{equation}
\begin{split}
S_{\rm sft}[\psi_s]&=\frac12\,\l\psi_s|Q|\psi_s\r=\frac12\,\int d^Dx\,\Big[\,\frac{1}{s!}\,\varphi^{\mu_1\ldots\mu_{s}}\B\varphi_{\mu_1\ldots\mu_{s}}-\frac{1}{(s-1)!}\,f^{\mu_1\ldots\mu_{s-1}}f_{\mu_1\ldots\mu_{s-1}}\\
&+\frac{2}{(s-1)!}\,f^{\mu_1\ldots\mu_{s-1}}\big(\del\cdot\varphi_{\mu_1\ldots\mu_{s-1}}-(s-1)\,\del_{\mu_1}\chi_{\mu_2\ldots\mu_{s-1}}\big)-\frac{1}{(s-2)!}\,\chi^{\mu_1\ldots\mu_{s-2}}\B\chi_{\mu_1\ldots\mu_{s-2}}
\Big]\;,    
\end{split}    
\end{equation}
assuming all fields to be real. The above action is automatically gauge invariant under $\delta\ket{\psi_s}=Q\ket{\Lambda_s}$,  since $Q^\dagger=Q$.
For $s=1$ the dilaton $\chi_{\mu_1\ldots\mu_{s-2}}$ is absent and one obtains
\begin{equation}
S=\int d^Dx\,\Big[\,\frac12\,A^\mu\B A_\mu-\frac12\,f^2+f\,\del\cdot A\Big]\;,  
\end{equation}
upon renaming $\varphi_\mu\to A_\mu$. Integrating out the auxiliary scalar $f$ one recovers the standard Maxwell action
\begin{equation}\label{Maxwell}
S=\int d^Dx\,\Big[\,\frac12\,A^\mu\B A_\mu+\frac12\,(\del\cdot A)^2\Big]=-\frac14\,\int d^Dx\,F^{\mu\nu}F_{\mu\nu}\;.  
\end{equation}

\subsection{$L_\infty$ interpretation}\label{sec:Loointerpret}

In this section we will interpret the BRST system discussed above as the $L_\infty$ chain complex of the spacetime field theory.
Homotopy Lie (or $L_\infty$) algebras \cite{Zwiebach:1992ie,Lada:1992wc,Hohm:2017pnh} encode the classical structure of perturbative gauge theories, in a similar way Lie algebras govern infinitesimal symmetries.
An $L_\infty$ algebra consists of an integer graded vector space $\cX=\bigoplus_iX_i$, endowed with multilinear brackets $B_n:\cX^{\otimes n}\rightarrow\cX$. These brackets obey a set of quadratic relations generalizing the Jacobi identity of Lie algebras. In the field theory context the $X_i$ represent the spaces of gauge parameters, fields, field equations and so on. The generalized Jacobi identities encode order by order the interactions, their consistency with gauge symmetries etc. 

To lowest order, an $L_\infty$ algebra consists of the graded vector space $\cX$ together with a nilpotent differential $B_1$ of degree $+1$. For a Lagrangian gauge theory the graded vector space $\cX$ typically consists of four subspaces, organized in the following chain complex:
\begin{equation}
\begin{tikzcd}[row sep=2mm]
X_{-1}\arrow{r}{B_1}&X_{0}\arrow{r}{B_1}&X_{1}\arrow{r}{B_1}&X_{2}\\
\Lambda&\psi&\cE&\cN\;,
\end{tikzcd}     
\end{equation}
where $X_{-1}$ is the space of gauge parameters $\Lambda$, $X_0$ the space of fields $\psi$, $X_1$ the space of field equations $\cE$ and $X_2$ the space of Noether identities $\cN$.
This organization is similar (in fact dual \cite{Jurco:2018sby,Arvanitakis:2020rrk,Grigoriev:2023lcc}) to the one of the Batalin-Vilkovisky formalism in terms of ghosts, fields and antifields.

Nilpotence of the differential expresses gauge invariance of the linearized field equations as $B_1^2(\Lambda)=0$, as well as the Noether identities between equations as $B_1^2(\psi)=0$. 
The worldline BRST system coincides with the $L_\infty$ chain complex $(\cX,B_1)$. We can in fact identify $\cX=\cH_s$ as the graded vector space of the $L_\infty$ algebra for spin $s$, with worldline ghost number as degree, i.e. $X_k=\cH_{s,k}$. Since $Q:\cH_{s,k}\rightarrow\cH_{s,k+1}$, and $Q^2=0$, we further identify the differential with the worldline BRST operator: $B_1=Q$. 

In agreement with the fact that symmetric tensors have irreducible gauge symmetries, the degree span for every value of $s$ (except $s=0$ of course) is $[-1,+2]$, yielding the chain complex
\begin{equation}
\begin{tikzcd}[row sep=2mm]
\cH_{s,-1}\arrow{r}{Q}&\cH_{s,0}\arrow{r}{Q}&\cH_{s,1}\arrow{r}{Q}&\cH_{s,2}\\
\Lambda_s&\psi_s&\cE_s&\cN_s\;.
\end{tikzcd}     
\end{equation}
The elements of the complex decompose according to the worldline ghost content as
\begin{equation}
\begin{split}
\Lambda_s&=\cB\,\xi_{s-1}\,\in\cH_{s,-1}\;,\\ \psi_s&=\varphi_s+c\cB\,f_{s-1}+\cC\cB\,\chi_{s-2}\,\in\cH_{s,0}\;,\\
\cE_s&=c\,E_s+\cC\,E_{s-1}+c\,\cC\cB\,E_{s-2}\,\in\cH_{s,1}\;,\\
\cN_s&=c\,\cC\,N_{s-1}\,\in\cH_{s,2}\;,
\end{split}    
\end{equation}
where we omitted the ket symbol and the component fields depend only on $x$ and $\alpha$'s, with their tensor rank indicated explicitly.
Here $\xi_{s-1}$, $\varphi_s$, $f_{s-1}$ and $\chi_{s-2}$ are the gauge parameter and triplet of fields introduced previously. $E_s, E_{s-1}$ and $E_{s-2}$ are the corresponding field equations, while $N_{s-1}$ is the single spin $s-1$ Noether identity, corresponding to the gauge parameter $\xi_{s-1}$.
The BRST operator $Q$ acts on objects of different degree as follows:
\begin{equation}
\begin{split}
Q\Lambda_s&=\del\xi_{s-1}+c\cB\,\B\xi_{s-1}+\cC\cB\,\del\cdot\xi_{s-1}\,\in\cH_{s,0}\;,\\
Q\psi_s&=c\,(\B\varphi_s-\del f_{s-1})+\cC\,(\del\cdot\varphi_s-\del\chi_{s-2}-f_{s-1})+c\,\cC\cB\,(\B\chi_{s-2}-\del \cdot f_{s-1})\,\in\cH_{s,1}\;,\\
Q\cE_s&=c\,\cC\,(\B E_{s-1}+\del E_{s-2}-\del\cdot E_{s})\,\in\cH_{s,2}\;,
\end{split}    
\end{equation}
where $\del$ denotes the symmetrized gradient and $\del\,\cdot$ the divergence.

The inner product on the Hilbert space $\cH$ is interpreted as an $L_\infty$ inner product in $\cX$. The fact that the basic overlap requires a $c$ ghost insertion (we remind that $\bra{0}c\ket{0}=1$) complies with the $L_\infty$ inner product having intrinsic degree $-1$ in our conventions.
This implies that gauge parameters $\Lambda_s$ in degree $-1$ are paired with Noether identities $\cN_s$ in degree $+2$, while fields $\psi_s$ in degree zero are paired with equations of motion $\cE_s$ in degree $+1$. Using the  overlap \eqref{overlap} together with the hermiticity assignments \eqref{ghostherm} and the vacuum condition \eqref{vacuum} we obtain
\begin{equation}
\begin{split}
\l\psi_s|\cE_s\r&=\l\cE_s|\psi_s\r\\
&=\int d^Dx\,\Big[\frac{1}{s!}\,\varphi^{\mu_1\ldots\mu_{s}}E_{\mu_1\ldots\mu_{s}}+\frac{1}{(s-1)!}\,f^{\mu_1\ldots\mu_{s-1}}E_{\mu_1\ldots\mu_{s-1}}-\frac{1}{(s-2)!}\,\chi^{\mu_1\ldots\mu_{s-2}}E_{\mu_1\ldots\mu_{s-2}}\Big]\;,\\
\l\Lambda_s|\cN_s\r&=\l\cN_s|\Lambda_s\r
=\frac{1}{(s-1)!}\int d^Dx\,\,\xi^{\mu_1\ldots\mu_{s-1}}N_{\mu_1\ldots\mu_{s-1}}\;,
\end{split}    
\end{equation}
where we assumed all fields to be real. Hermiticity of the BRST operator
$Q^\dagger=Q$ coincides at this order with the $L_\infty$ algebra being cyclic, which ensures that the corresponding field theory admits an action principle.

Although the worldline theory describes particles of all spins, our primary interest is in describing Yang-Mills theory in first-quantized form. In the following we will thus restrict to the $s=1$ sector of the theory, associated to the Hilbert subspace $\cH_1$.

\section{Spin one particle in Yang-Mills background}\label{sec:background coupling}

The worldline model so far is a free theory that describes a single spin one particle in the $s=1$ sector. In order to introduce interactions we will couple the worldline to a background Yang-Mills field by deforming the BRST operator. To this end one has to first add color degrees of freedom to the particle, to which we turn next.

\subsection{Color degrees of freedom}

Our goal is to extend the worldline Hilbert space $\cH$ so as to accommodate representations of a color Lie algebra $\mathfrak{g}
$, which we take to be compact and semisimple. To this end, we introduce a conjugate pair of worldline fields $w_a(\tau)$ and $\bar w^a(\tau)$ with action \cite{Bastianelli:2013pta,Bastianelli:2015iba}
\begin{equation}
S_{\rm color}=\int d\tau \big[-i\bar w^a\dot w_a\big] \;, 
\end{equation}
where $a,b=1,\ldots,{\rm dim}\mathfrak{g}$ are adjoint indices of $\mathfrak{g}$. We take the Killing form to be $\kappa_{ab}=-\delta_{ab}$ and use $\delta_{ab}$ and its inverse to lower and raise indices, so that we can impose the reality condition $(w_a)^*=\bar w_a$.

Upon canonical quantization the color vectors obey the creation-annihilation algebra
\begin{equation}
[\bar w^a,w_b]=\delta^a{}_b\;.    
\end{equation}
We can thus construct the associated Hilbert space $\cH_{\rm color}$ as the Fock space of creation operators $w_a$ acting on a vacuum $\ket{0}_{\rm color}$ annihilated by $\bar w^a$. The inner product on $\cH_{\rm color}$ is given by declaring $w_a^\dagger=\bar w_a$. The resulting space is the direct sum of symmetrized products of the adjoint representation of $\mathfrak{g}$: $\cH_{\rm color}=\bigoplus_{r=0}^\infty\cH_{\rm color}^r$. A generic vector is given by
\begin{equation}
\ket{V}_{\rm color} =\sum_{r=0}^\infty\frac{1}{r!}V^{a_1\cdots a_r}\,w_{a_1}\cdots w_{a_r}\ket{0}_{\rm color} \;,  
\end{equation}
where the tensor rank $r$ is counted by the number operator $N_w=w_a\bar w^a$.
We can use the structure constants $f_{ab}{}^c$ to define the generators of $\mathfrak{g}$ acting on these representations as
\begin{equation}\label{T generators}
T_a:=f_{ab}{}^c\,w_c\bar w^b\;\qquad \longrightarrow\qquad[T_a,T_b]=f_{ab}{}^c\,T_c\;,\quad T_a^\dagger=-T_a\;. 
\end{equation}

From now on we will restrict ourselves to the adjoint representation $\cH_{\rm color}^1$, which is the eigenspace with $N_w=1$. The monomials $\ket{w_a}=w_a\ket{0}_{\rm color}$ form a basis of $\cH_{\rm color}^1$ and the identity decomposes as $\mathds{1}=\ket{w_a}\bra{\bar w^a}    
$. The inner product between two adjoint elements involves the metric $\delta_{ab}$ as
\begin{equation}
\l U|V\r=U^aV^b\,\l \bar w_a|w_b\r=\delta_{ab}\,U^aV^b \;.   
\end{equation}
The standard definition of the Killing form as a trace over the adjoint representation can be obtained upon using the identity decomposition:
\begin{equation}\label{tr}
{\rm tr}\big(T_a T_b\big)=\bra{\bar w^c}T_aT_b\ket{w_c}=f_{ac}{}^df_{bd}{}^c =\delta_{ab}\;.   
\end{equation}

\subsection{Deformed BRST charge}

Upon adding the color sector, the full Hilbert space of the worldline theory is given by the tensor product $\cH\otimes\cH_{\rm color}$. Since the BRST operator $Q$  is diagonal in spin and acts trivially on $\cH_{\rm color}$, we restrict to the spin one sector in $\cH$ and to the adjoint representation in $\cH_{\rm color}$, thereby working on the graded vector space
\begin{equation}
\cX:=\cH_1\otimes\cH_{\rm color}^1\;,    
\end{equation}
with the degree given by the worldline ghost number as discussed previously. All elements of $\cX$ (corresponding to gauge parameters, fields etc.) are valued in the adjoint representation of $\mathfrak{g}$. For instance, a field in degree zero can be expanded as
\begin{equation}\label{gluon}
\ket{\psi}=\Big(a^a_\mu(x)\,\alpha^\mu\ket{0}+f^a(x)\,c\,\cB\ket{0}\Big)\otimes \ket{w_a}\;,
\end{equation}
where $f^a$ is the auxiliary scalar field. Here we use a lower case $a_\mu$ for the gluon state, as we will reserve capital $A_\mu$ for the background gauge field deforming the BRST operator.

To this end, we rewrite $Q$ as
\begin{equation}\label{ghost combos}
\begin{split}
Q&=c\,\B+S^\mu\,\del_\mu-\cM\,b\;,\\
S^\mu&:=\bar\cC\,\alpha^\mu+\cC\,\bar\alpha^\mu\;,\\
\cM&:=\cC\,\bar\cC\;,
\end{split}   
\end{equation}
where we kept explicit the $b,c$ ghosts and spacetime derivatives of the various terms. We further introduce the Lorentz spin generator, which rotates the $\alpha^\mu$ oscillators: 
\begin{equation}
\begin{split}
S^{\mu\nu}&:=\alpha^\mu\bar\alpha^\nu-\alpha^\nu\bar\alpha^\mu\;,\\    
[S^{\mu\nu},S^\rho]&=2\,\eta^{\rho[\nu}S^{\mu]}\;,\quad
[S^{\mu\nu},S^{\rho\sigma}]=4\,\eta^{[\rho[\nu}S^{\mu]\sigma]}\;.  
\end{split}
\end{equation}
The ghost vector $S^\mu$, Lorentz generator $S^{\mu\nu}$ and $\cM$ all commute with the $U(1)$ generator $\cJ$ and obey
\begin{equation}
\begin{split}
S^\mu S^\nu&=\cM\,(\eta^{\mu\nu}-S^{\mu\nu})\;,\quad S^\mu\cM=\cM S^\mu=0 \;,\\
[S^{\mu\nu},Q]&=2\,S^{[\mu}\del^{\nu]}\;.
\end{split}    
\end{equation}

We now introduce the background gauge field and the corresponding covariant derivative as quantum mechanical operators acting on the Hilbert space $\cX$:
\begin{equation}\label{Dmu}
\cA_\mu:=A_\mu^a\,T_a=A_\mu^a(x)\,f_{ab}{}^c\,w_c\bar w^b\;,\qquad \cD_\mu:=\del_\mu+\cA_\mu\;.    
\end{equation}
As such, the ordinary covariant derivative $D_\mu$ is produced by the left action of $\cD_\mu$ on states and by its commutator on operators. For instance, given a gauge parameter $\ket{\Lambda}=\lambda^a(x)\,\cB\ket{0}\otimes\ket{w_a}$ one has $\cD_\mu\ket{\Lambda}=D_\mu\lambda^a\,\cB\ket{0}\otimes\ket{w_a}$, while for an operator $\Lambda=\lambda^a(x)\,T_a$ the covariant derivative is given by $[\cD_\mu,\Lambda]=D_\mu\lambda^a\,T_a$.
Taking this into account, the operator $\cD_\mu$ obeys
\begin{equation}
\begin{split}
[\cD_\mu,\cD_\nu]&=\cF_{\mu\nu}\;,\quad\hspace{3mm}\cF_{\mu\nu}:=\del_\mu\cA_\nu-\del_\nu\cA_\mu+[\cA_\mu,\cA_\nu]\;,\\
\cF_{\mu\nu}&=F_{\mu\nu}^a\,T_a\;,\quad F_{\mu\nu}^a=\del_\mu A^a_\nu-\del_\nu A^a_\mu+f_{bc}{}^aA^b_\mu A^c_\nu\;,  
\end{split}    
\end{equation}
where the bracket above is the quantum mechanical commutator.
We define the deformed BRST operator $Q_A$ by replacing $\del_\mu\rightarrow\cD_\mu$ in \eqref{ghost combos} and adding a non-minimal coupling to $\cF_{\mu\nu}$:
\begin{equation}\label{QA}
Q_A:=c\,\triangle+S^\mu\, \cD_\mu-\cM\,b \;,\qquad  \triangle:=\cD^\mu \cD_\mu+\cF_{\mu\nu}\,S^{\mu\nu}\;.
\end{equation}

For the worldline theory to be quantum mechanically consistent (which requires the decoupling of unphysical states), we demand that $Q_A^2=0$. Computing the square one obtains
\begin{equation}\label{QAsquare}
Q_A^2=-\frac32\,\cM\,S^{\mu\nu}\cF_{\mu\nu}-c\,\Big(S^\mu\,D^\nu\cF_{\mu\nu}+D_\mu\cF_{\nu\rho}\,S^\mu S^{\nu\rho}\Big) \;,   
\end{equation}
where we denoted the operator corresponding to the covariant derivative of $F_{\mu\nu}$ by 
\begin{equation}
D_\mu\cF_{\nu\rho}:=[\cD_\mu,\cF_{\nu\rho}]=D_\mu F_{\nu\rho}^a\,T_a\;.  
\end{equation}
As one can see explicitly, the deformed BRST operator is not nilpotent unless $\cF_{\mu\nu}=0$. However, \eqref{QAsquare} is an operator equation holding on the full Hilbert space $\cH\otimes\cH_{\rm color}$. Physically, this expresses the fact that higher spin fields do not admit minimal coupling to Yang-Mills. If we restrict $Q_A$ to act on $\cX$ (which, in particular, has occupation number $\cJ=1$), $\cM\,S^{\mu\nu}\rvert_{\cX}=0$, since it has two annihilation operators on the right. Similarly, we can rewrite the last term in normal ordering and restrict it to $\cX$:
\begin{equation}
\begin{split}
S^\mu\,S^{\nu\rho}\rvert_{\cX}&=2\,(\bar\cC\,\alpha^\mu+\cC\,\bar\alpha^\mu)\,\alpha^{[\nu}\bar\alpha^{\rho]}\rvert_{\cX}\\
&=2\,\big(\alpha^\mu\alpha^{[\nu}\bar\alpha^{\rho]}\bar\cC+\cC\,\alpha^{[\nu}\bar\alpha^{\rho]}\bar\alpha^\mu+\cC\,\eta^{\mu[\nu}\bar\alpha^{\rho]}\big)\rvert_{\cX}\\
&=2\,\cC\,\eta^{\mu[\nu}\bar\alpha^{\rho]}\;,
\end{split}    
\end{equation}
where we discarded any term with two annihilation operators on the right, which give zero on any state in $\cX$.
When restricting $Q_A$ to $\cX$ we thus find
\begin{equation}
Q_A^2\rvert_{\cX}=c\,(\bar\cC\,\alpha^\mu-\cC\,\bar\alpha^\mu)\,D^\rho\cF_{\rho\mu}\;.
\end{equation}
We see that the condition for nilpotence of $Q_A$ is the field equation for the background $A_\mu$, which was also found in \cite{Dai:2008bh} for the case of the $\cN=2$ supersymmetric worldline. This feature, which sometimes is viewed as magical in string theory, has  a natural interpretation once we combine the first-quantized and field theoretic perspectives. As an aside, notice that if we restrict to the subspace with $\cJ=0$, which contains only a scalar field, $Q_A$ is nilpotent without any condition on the background, as expected from scalar QCD.

\subsection{Spacetime interpretation}

In order to see why $Q_A$ is nilpotent only when the background is on-shell, let us consider the spacetime action for the gluon fluctuation \eqref{gluon} in the presence of the $A_\mu$ background:
\begin{equation}\label{S BG}
\begin{split}
S_{{\rm sft},A}[\psi]&=\frac12\,\bra{\psi}Q_A\ket{\psi}=\int d^Dx\,\Big[-\frac12\,D^\mu a_a^\nu D_\mu a^a_\nu-\frac12\,f_af^a+f_a\,D^\mu a^a_\mu-f_{bc}{}^a\,F_a^{\mu\nu}a^b_\mu a^c_\nu\Big]\;,
\end{split}    
\end{equation}
where $D_\mu a_\nu^a=\del_\mu a_\nu^a+f_{bc}{}^a\,A_\mu^ba_\nu^c$ and we raise and lower color indices with $\delta_{ab}$. The above action is invariant under the deformed gauge transformation $\delta\ket{\psi}=Q_A\ket{\Lambda}$ if and only if $Q_A^2=0$. Gauge invariance of \eqref{S BG} ensures that the unphysical polarizations of the gluon $a_\mu$ decouple, which is equivalent to the consistency of the worldline quantum theory.

To proceed further we integrate out the auxiliary field $f^a$, thus obtaining
\begin{equation}\label{S2YM}
S_{{\rm sft},A}[a]=\int d^Dx\,\Big[-\frac14\,(D^\mu a_a^\nu-D^\nu a_a^\mu)(D_\mu a^a_\nu-D_\nu a^a_\mu)-\frac12\,f_{bc}{}^a\,F_a^{\mu\nu}a^b_\mu a^c_\nu\Big]   \;. \end{equation}
This is nothing but the Yang-Mills action for ${\bf A}_\mu=A_\mu+a_\mu$ at quadratic order in $a_\mu$. To establish the connection with the field equation of $A_\mu$, we take the full Yang-Mills action for ${\bf A_\mu}$ and expand it in powers of the fluctuation:
\begin{equation}\label{SBGexp}
\begin{split}
S_{\rm YM}[{\bf A}]&=S_{\rm YM}[A]+S_1[A;a]+S_2[A;a]+\cO(a^3)\;,\\
S_1[A;a]&=\int d^Dx\,a_\mu^a\left.\frac{\delta S_{\rm YM}}{\delta{\bf A}_\mu^a}\right\rvert_{{\bf A}=A}=\int d^Dx\,(D^\mu F_{\mu\nu}^a)\,a^\nu_a\;,\\ 
S_2[A;a]&=S_{{\rm sft},A}[a]\;,
\end{split}    
\end{equation}
where $S_k[A;a]$ contains $k$ powers of $a_\mu$. The action $S_{\rm YM}[{\bf A}]$ is clearly gauge invariant under $\delta{\bf A}^a_\mu={\bf D}_\mu\lambda^a=\del_\mu\lambda^a+f_{bc}{}^a{\bf A}_\mu^b\lambda^c$. In the background field expansion with ${\bf A}_\mu=A_\mu+a_\mu$ this is the same as keeping $A_\mu$ fixed and transforming the fluctuation as
\begin{equation}\label{gaugeBGexp}
\delta a_\mu^a=D_\mu\lambda^a+f_{bc}{}^a a_\mu^b\lambda^c=\delta_0 a_\mu^a+\delta_1 a_\mu^a\;,    
\end{equation}
with the subscript on the variation counting again the powers of $a_\mu$. Gauge invariance of the action \eqref{SBGexp} under \eqref{gaugeBGexp} gives relations order by order in powers of $a_\mu$. The zeroth order in $a_\mu$ is the Noether identity for the background: $D^\mu D^\nu F_{\mu\nu}^a\equiv0$, while to linear order we obtain
\begin{equation}
\delta_0S_2[A;a]+\delta_1S_1[A;a]=0  \;.  
\end{equation}
This means that the quadratic action $S_2[A;a]\equiv S_{{\rm sft},A}[a]$ is gauge invariant under $\delta_0 a_\mu^a=D_\mu\lambda^a$ only if the background $A_\mu$ is on-shell, since then $S_1[A;a]=0$. On the worldline Hilbert space the variation $\delta_0$ is given by $Q_A\ket{\Lambda}$, which explains why $Q_A^2=0$ only if the background satisfies the field equations.

This discussion should make it clear that the Hilbert space $\cX$ together with the BRST operator $Q_A$ contain information on the full Yang-Mills action via \eqref{S2YM}. More than that, it turns out that $Q_A$ alone already captures the full nonlinear structure of Yang-Mills, including gauge transformations and Noether identities, as we will establish in the next section.

\section{Off-shell vertex operators and nonlinear theory}\label{sec:vertex operators}

In this section we focus on the algebra of operators acting on the Hilbert space $\cX$.
Associating the gauge field $A_\mu$ to the BRST operator $Q_A$, we will show that the entire nonlinear structure of Yang-Mills theory, encoded in its $L_\infty$ algebra, is contained in the algebra of vertex operators acting on $\cX$.

\subsection{Maurer-Cartan equation and vertex operators}

We start from the deformed BRST operator $Q_A$ as in \eqref{QA}:
\begin{equation}
Q_A=c\,\big(\cD^\mu \cD_\mu+\cF_{\mu\nu}\,S^{\mu\nu}\big)+S^\mu\, \cD_\mu-\cM\,b \;,
\end{equation}
which we view as a map that takes the gauge field $A_\mu$ and produces an operator acting on $\cX$. Since $Q_A$ is not linear in the gauge field, it defines two types of vertex operators upon expanding it in powers of $A_\mu$:
\begin{equation}\label{Qexpanded}
\begin{split}
Q_A&=Q+\cV(A)+\tfrac12\,\cV_2(A,A)\;,\\
\cV(A)&:=S^\mu\cA_\mu+c\,\Big(2\,\cA^\mu\del_\mu+(\del^\mu\cA_\mu)+2\,(\del_\mu\cA_\nu)\,S^{\mu\nu}\Big)\;,\\
\cV_2(A,A)&:=2\,c\,\big(\cA^2+[\cA^\mu,\cA^\nu]\,S_{\mu\nu}\big)\;,
\end{split}    
\end{equation}
where we recall that $\cA_\mu=A_\mu^a\,T_a$. Here $\cV(A)$ is the usual linear vertex operator, while $\cV_2(A,A)$ is a bilinear vertex whose role will become clear in the following.

In the previous section we have shown that $Q_A^2$, when restricted to $\cX$, is proportional to the Yang-Mills field equation. In the following we will always restrict the products of operators\footnote{The restricted product remains associative, since $\big(\cO_1\rvert_\cX\cO_2\rvert_\cX\big)\rvert_\cX=\big(\cO_1\cO_2\big)\rvert_\cX$ for operators of $U(1)$ charge zero, which commute with the occupation number $\cJ$.} to act on $\cX$, but for notational simplicity we will omit the restriction symbol $\rvert_\cX$. Given the expansion \eqref{Qexpanded}, we interpret $Q_A^2$ as a generalized Maurer-Cartan equation for the vertex operators $\cV(A)$ and $\cV_2(A,A)$:
\begin{equation}
Q_A^2=\{Q,\cV(A)\}+\tfrac12\,\{\cV(A),\cV(A)\}+\tfrac12\,\{Q,\cV_2(A,A)\}+\tfrac12\,\{\cV(A),\cV_2(A,A)\}\;.    
\end{equation}
Since $Q_A^2=c\,(\bar\cC\,\alpha^\mu-\cC\,\bar\alpha^\mu)\, D^\rho F^a_{\rho\mu}\,T_a$, the Maurer-Cartan equation for the vertex operators is in one-to-one correspondence with the perturbative expansion of the field equation for $A_\mu$.
Moreover, $Q_A^2$ defines a linear vertex operator for the field equation $E_\mu^a=D^\rho F_{\rho\mu}^a$, which is mapped to the ghost number two operator
\begin{equation}
\cV(E):= c\,\tilde S^\mu\cE_\mu \;,\quad \tilde S^\mu:=\bar\cC\,\alpha^\mu-\cC\,\bar\alpha^\mu\;,\quad\cE_\mu=E_\mu^a\,T_a\;.  
\end{equation}
Notice that by linear vertex operator we mean that $\cV(E)$ is linear in the equation of motion and does not contain extra powers of the field.

The Maurer-Cartan equation is covariant by construction under the operator gauge transformation
\begin{equation}
\delta Q_A=[Q_A,\cV(\lambda)]\;\longrightarrow\;\delta(Q_A^2)=[Q_A^2,\cV(\lambda)]\;, 
\end{equation}
where in general $\cV(\lambda)$ can be any ghost number zero operator commuting with $\cJ$.
It turns out that the simplest choice for $\cV(\lambda)$ is the one that reproduces the Yang-Mills gauge symmetry. Upon taking
$\cV(\lambda)=\lambda^a\,T_a$, the commutator is given by
\begin{equation}
[Q_A,\cV(\lambda)]=c\,\Big(2\,(D^\mu\Lambda)\, \cD_\mu+(D^2\Lambda)+[\cF_{\mu\nu},\Lambda]\,S^{\mu\nu}\Big)+S^\mu (D_\mu\Lambda)\;, \end{equation}
where $\Lambda=\lambda^a\,T_a$.
This coincides with varying $Q_A$ by taking a variation of $A_\mu$, meaning that
\begin{equation}\label{delta Q}
[Q_A,\cV(\lambda)]=Q_{A+\delta_\lambda A}-Q_A\;,    
\end{equation}
keeping only the first order in $\delta A_\mu$, as it fits a variation. Notice that, since $Q_A$ is \emph{not} linear in $A_\mu$, $\delta Q_A\neq Q_{\delta A}$. This has important consequences that we will elucidate in the next section. 

Finally, given that $Q_A^2$ yields the vertex operator $\cV(E)$ for the field equation, the vertex operator for the Noether identity $\cN=N^a\,T_a$
must follow from
\begin{equation}
[Q_A,\cV(E)]=-c\,\cM\,D^\mu E_\mu^a\,T_a\;,\quad\cV(N):=c\,\cM \,\cN\;,   
\end{equation}
since it vanishes identically for $\cV(E)=Q_A^2$. We summarize here the linear vertex operators:
\begin{equation}\label{LooVOs}
\begin{array}{lcc}
\cV(\lambda)=\Lambda\;,&|\cV(\lambda)|=0\;,& \Lambda=\lambda^a\,T_a\;, \\
\cV(A)=S^\mu\cA_\mu+c\,\Big(2\,\cA^\mu\del_\mu+(\del\cdot\cA)+2\,(\del_\mu\cA_\nu)\,S^{\mu\nu}\Big)\;,&|\cV(A)|=1\;,&\cA_\mu=A^a_\mu\,T_a\;,\\
\cV(E)=c\,\tilde S^\mu\,\cE_\mu\;,& |\cV(E)|=2\;,&\cE_\mu=E_\mu^a\,T_a\;,\\
\cV(N)=c\,\cM\,\cN\;,& |\cV(N)|=3\;,&\cN=N^a\,T_a\;,
\end{array}    
\end{equation}
with the degree given by ghost number. There is a single bilinear vertex for two fields, obtained by symmetrizing $\cV_2(A,A)$ in the two inputs: 
\begin{equation}\label{V2}
\cV_2(A_1,A_2)=c\,\big(\cA_{1}\cdot\cA_{2}+\cA_{2}\cdot\cA_{1}+2\,[\cA_{1}^\mu,\cA_{2}^\nu]\,S_{\mu\nu}\big)\;.    
\end{equation}
All vertex operators commute with the $U(1)$ generator $\cJ$. This ensures that they are well defined on $\cX$, meaning that their products can be restricted to $\cX$ consistently.

It may look unfamiliar to assign vertex operators for gauge parameters and field equations. If we worked with the Batalin-Vilkovisky formalism these would be vertex operators for ghosts and antifields.
The operator Maurer-Cartan equation and its gauge symmetries guarantee that the entire $L_\infty$ algebra of Yang-Mills is encoded in these operator relations. In the following we will see precisely how it is embedded.

\subsection{Vertex operators and $L_\infty$ algebra}

In order to formalize the perturbative expansion of the Yang-Mills equations and gauge transformations, we need some basic definitions about $L_\infty$ algebras. As mentioned in section \ref{sec:Loointerpret}, an $L_\infty$ algebra consists of a graded vector space endowed with multilinear brackets $B_n$, obeying some generalized Jacobi identities. We use conventions where all brackets $B_n$ have intrinsic degree $+1$ and are graded symmetric with respect to the $L_\infty$ degree.

For the case of Yang-Mills theory, the graded vector space $\cX^{\rm YM}$ contains gauge parameters, fields, field equations and Noether identities, organized in the following chain complex:
\begin{equation}\label{YM complex}
\begin{tikzcd}[row sep=2mm]
X^{\rm YM}_{-1}\arrow{r}{B_1}&X^{\rm YM}_{0}\arrow{r}{B_1}&X^{\rm YM}_{1}\arrow{r}{B_1}&X^{\rm YM}_{2}\\
\lambda^a&A_\mu^a&E_\mu^a&N^a\;,
\end{tikzcd}     
\end{equation}
with the subscript in $X^{\rm YM}_{k}$ denoting the $L_\infty$ degree. The differential $B_1$ has degree $+1$ and is given by
\begin{equation}
(B_1\lambda)_\mu^a=\del_\mu\lambda^a\;,\quad (B_1A)_\mu^a=\B A_\mu^a-\del_\mu\del\cdot A^a \;,\quad (B_1E)^a=-\del^\mu E_\mu^a \;,
\end{equation}
thus describing the linearized gauge transformation, equation of motion and Noether identity, respectively. 
Coming to the nonlinear structure, for now we recall that the expansion of field equations and gauge transformations in powers of the field defines the brackets $B_n(A_1,\ldots,A_n)$ and $B_n(\lambda,A_1,\ldots,A_{n-1})$
via
\begin{equation}
\begin{split}
D^\rho F_{\rho\mu}^a&=\left(B_1(A)+\tfrac12\,B_2(A,A)+\tfrac{1}{3!}\,B_3(A,A,A)\right)_\mu^a \,\in X_1^{\rm YM}\;,\\
\delta_\lambda A_\mu^a&=D_\mu\lambda^a=\left(B_1(\lambda)+B_2(\lambda,A)\right)_\mu^a \,\in X_0^{\rm YM}\;.
\end{split}    
\end{equation}

We now use the above definition of the brackets to compare the expansion of $Q_A^2$ in terms of vertex operators with the expansion of $c\,\tilde S^\mu\,D^\rho F_{\rho\mu}^a\,T_a$ in powers of the field:
\begin{equation}
\begin{split}
Q_A^2&=\{Q,\cV(A)\}+\tfrac12\,\{\cV(A),\cV(A)\}+\tfrac12\,\{Q,\cV_2(A,A)\}+\tfrac12\,\{\cV(A),\cV_2(A,A)\}\\
&=c\,\tilde S^\mu\,D^\rho F_{\rho\mu}^a\,T_a=\cV\Big(B_1(A)+\tfrac12\,B_2(A,A)+\tfrac{1}{3!}\,B_3(A,A,A)\Big)\;.
\end{split}    
\end{equation}
Matching both sides order by order in the gauge field we derive the following relations for the vertex operators of the $L_\infty$ brackets:
\begin{equation}\label{Bas}
\begin{split}
\cV\big(B_1(A)\big)&=\{Q,\cV(A)\}\\
\cV\big(B_2(A,A)\big)&=\{\cV(A),\cV(A)\}+\{Q,\cV_2(A,A)\}\\
\cV\big(B_3(A,A,A)\big)&=3\,\{\cV(A),\cV_2(A,A)\}\;.
\end{split}    
\end{equation}
This shows quite clearly that the non-vanishing $\cV_2(A,A)$ is responsible for the presence of higher brackets in the $L_\infty$ algebra. Moreover, the three-bracket $B_3(A,A,A)$ is derived by combining at most bilinear operators.

We now use the same strategy to identify how the brackets for gauge transformations are embedded in the vertex operators. We expand the operator relation $\delta Q_A=[Q_A,\cV(\lambda)]$ and use \eqref{delta Q} to find
\begin{equation}
\begin{split}
\delta Q_A&=[Q,\cV(\lambda)]+[\cV(A),\cV(\lambda)]+\tfrac12\,[\cV_2(A,A),\cV(\lambda)]\\
&=\cV(\delta A)+\cV_2(\delta A,A)=\cV\Big(B_1(\lambda)+B_2(\lambda,A)\Big)+\cV_2\Big(B_1(\lambda)+B_2(\lambda,A),A\Big)\;,
\end{split}       
\end{equation}
upon taking into account that $\cV_2(A_1,A_2)$ is symmetric in the two inputs.
Matching the two expressions order by order in $A_\mu$ we identify the vertex operators for the following brackets:
\begin{equation}\label{Blambdas}
\begin{split}
\cV\big(B_1\lambda\big)&=[Q,\cV(\lambda)]\;,\\
\cV\big(B_2(\lambda,A)\big)&=-[\cV(\lambda),\cV(A)]-\cV_2\big(B_1\lambda,A\big)\;,\\
\cV_2\big(B_2(\lambda,A),A\big)&=-\tfrac12\,[\cV(\lambda),\cV_2(A,A)]\;.        
\end{split}    
\end{equation}

The remaining brackets of the $L_\infty$ algebra are similarly related to commutators of vertex operators. The relations are determined by following the same procedure for the closure of gauge transformations, gauge covariance of the field equations and so on, yielding
\begin{equation}\label{remaining B2s}
\begin{split}
\cV\big(B_1(E)\big)&=[Q,\cV(E)]\;,\\
\cV\big(B_2(\lambda_1,\lambda_2)\big)&=-[\cV(\lambda_1),\cV(\lambda_2)]\;,\\
\cV\big(B_2(\lambda,E)\big)&=-[\cV(\lambda),\cV(E)]\;,\\
\cV\big(B_2(A,E)\big)&=[\cV(A),\cV(E)]\;,\\
\cV\big(B_2(\lambda,N)\big)&=-[\cV(\lambda),\cV(N)]\;,
\end{split}    
\end{equation}
following the sign conventions of \cite{Bonezzi:2022yuh}. All further commutators of $\cV$ and $\cV_2$ are trivial:
\begin{equation}\label{V2V}
[\cV_2(A_1,A_2),\cV(E)]=0\;,\quad [\cV_2(A_1,A_2),\cV(N)]=0\;,
\end{equation}
which agree with the absence of bilinear vertices $\cV_2$ and three-brackets $B_3$ other than $\cV_2(A_1,A_2)$ and $B_3(A_1,A_2,A_3)$.
This exhausts all the non-vanishing brackets of the $L_\infty$ algebra of Yang-Mills.
The two-brackets in \eqref{remaining B2s} have also a familiar interpretation in gauge theory: $B_2(\lambda_1,\lambda_2)$ encodes the algebra of gauge transformations, $B_2(\lambda,E)$ and $B_2(\lambda,N)$ express covariance of the field equations and Noether identity, respectively, while $B_2(A,E)$ is the nonlinear contribution to the Noether identity $D^\mu E_\mu^a=0$.

We can summarize the above dictionary between $L_\infty$ brackets and vertex operators in a unified fashion. To this end, we shall denote generic elements of the Yang-Mills complex \eqref{YM complex} as $X,Y,Z,\ldots\in\cX^{\rm YM}$. We further introduce the graded commutator of operators, defined as
\begin{equation}
[O_1,O_2\}= O_1O_2-(-1)^{|O_1||O_2|}O_2O_1\;,   
\end{equation}
with the degree $|O_i|$ given by ghost number. Comparing \eqref{LooVOs} with \eqref{YM complex}, the $L_\infty$ degree of elements of $\cX^{\rm YM}$ is related to the degree of their vertex operators by
\begin{equation}
|\cV(X)|=|X|+1\;,\quad |\cV_2(X,Y)|=|X|+|Y|+1\;.   
\end{equation}
Here we write a general bilinear vertex $\cV_2(X,Y)$, which we define to be graded symmetric: $\cV_2(X,Y)=(-1)^{|X||Y|}\cV_2(Y,X)$. 
In this specific case $\cV_2$ is non-vanishing only when both $X$ and $Y$ are fields, i.e. elements of $X^{\rm YM}_0$, and is given by \eqref{V2}.
The vertex operators for the brackets $B_n$ are then given by
\begin{equation}\label{Vbrackets dictionary}
\begin{split}
\cV\big(B_1(X)\big)&=[Q,\cV(X)\}\;,\\
\cV\big(B_2(X,Y)\big)&=(-1)^{|X|}[\cV(X),\cV(Y)\}\\
&+[Q,\cV_2(X,Y)\}-\cV_2\big(B_1(X),Y\big)-(-1)^{|X|}\cV_2\big(X,B_1(Y)\big)\;,\\
\cV\big(B_3(X,Y,Z)\big)&=[\cV(X),\cV_2(Y,Z)\}+\cV_2\big(B_2(X,Y),Z\big)+\;\text{graded cyclic}\;,
\end{split}    
\end{equation}
where the sign for graded cyclic permutations is given by moving inputs past one another, such as $(X,Y,Z)\rightarrow(-1)^{|X|(|Y|+|Z|)}(Y,Z,X)$. Taking $\cV_2$ to be non-vanishing only for two fields, one recovers the relations \eqref{Bas}, \eqref{Blambdas} and \eqref{remaining B2s} upon specifying the degrees of the inputs.

The $L_\infty$ algebra of Yang-Mills theory has non-vanishing brackets up to a single $B_3$. In this case, the generalized Jacobi identities (which can be infinitely many in general) reduce to
\begin{equation}\label{genJac}
\begin{split}
B_1^2(X)&=0\;,\\
B_1\big(B_2(X,Y)\big)+B_2\big(B_1(X),Y\big)+(-1)^{|X|}B_2\big(X,B_1(Y)\big)&=0\;,\\
\Big(B_2\big(B_2(X,Y),Z\big)+B_3\big(B_1(X),Y,Z\big)+\;\text{graded cyclic}\Big)+B_1\big(B_3(X,Y,Z)\big)&=0\;.
\end{split}    
\end{equation}
These express that $B_1$ is a nilpotent differential acting as a derivation on $B_2$, while $B_2$ obeys the graded Jacobi identity up to homotopy, given by $B_3$. From a field theory perspective, the above relations encode the usual consistency conditions order by order in perturbation theory.
Given the vertex operators \eqref{Vbrackets dictionary} for the brackets, the generalized Jacobi identities \eqref{genJac} follow, thanks to the fact that operators form a graded Lie algebra with respect to graded commutators.

\subsection{Spacetime action from vertex operators}

Having established the relation between vertex operators and the $L_\infty$ algebra of Yang-Mills, in this last section we will show that the spacetime action is obtained as an expectation value of off-shell vertex operators.

To this end, we first introduce a ``physical'' vacuum state, which we denote as $\ket{1}$, by acting on the Fock vacuum with the antighost creation operator $\cB$: 
\begin{equation}
\ket{1}:=\cB\ket{0} \;,\quad \bra{1}:=\bra{0}\bar\cB=-(\ket{1})^\dagger
\end{equation}
Contrary to the Fock vacuum, the state $\ket{1}$ has $\cJ=1$. It thus belongs to the space $\cH_1$ and obeys $Q\ket{1}=0$. Upon tensoring with the color basis, the states $\ket{1}\otimes\ket{w_a}$ belong to the space $\cX$ and are physical in the sense that they coincide with constant gauge parameters.
Given a local operator $\cO(x)$ acting on $\cX=\cH_1\otimes\cH_{\rm color}^1$, we define its vacuum expectation value by
\begin{equation}\label{vev}
\big\l\cO(x)\big\r:=\int d^Dx\,{\rm tr}\,\bra{1}\cO(x)\ket{1}=\int d^Dx\,\bra{\bar w^a}\otimes\bra{1}\cO(x)\ket{1}\otimes\ket{w_a}\;,    
\end{equation}
where we take the trace over the color degrees of freedom using \eqref{tr}.

Acting with the vertex operators $\cV(X)$ on $\ket{1}$ (but not yet on $\ket{w_a}$) we obtain states in $\cH_1$ times generators $T_a$, which are still operators on $\cH_{\rm color}$:
\begin{equation}
\begin{split}
\cV(\lambda)\ket{1}&=\lambda^a(x)\,\cB\ket{0}\otimes T_a \;,\\
\cV(A)\ket{1}&=\big(A_\mu^a(x)\,\alpha^\mu+\del^\mu A_\mu^a(x)\,c\,\cB\big)\ket{0}\otimes T_a\;,\\
\cV(E)\ket{1}&=E_\mu^a(x)\,c\,\alpha^\mu\ket{0}\otimes T_a\;,\\
\cV(N)\ket{1}&=N^a(x)\,c\,\cC\ket{0}\otimes T_a \;.
\end{split}    
\end{equation}
This gives a sort of operator-state correspondence between vertex operators $\cV(X)$ and the Yang-Mills graded vector space $\cX^{\rm YM}$. Using the expectation value \eqref{vev}, the standard $L_\infty$ pairing between fields and field equations in $\cX^{\rm YM}$ is given by
\begin{equation}\label{vevinner}
\big\l\cV(A)\cV(E)\big\r=\int d^Dx\,A_\mu^a(x)\,E^\mu_a(x)\;.   
\end{equation}
Since the $L_\infty$ algebra of Yang-Mills is cyclic (which is guaranteed, being a Lagrangian theory), the action can be written in the generalized Maurer-Cartan form
\begin{equation}
S_{\rm YM}[A] =\int d^Dx\,A_\mu^a\,\big[\tfrac12\,B_1(A)+\tfrac{1}{3!}\,B_2(A,A)+\tfrac{1}{4!}\,B_3(A,A,A)\big]^\mu_a    \;.
\end{equation}
Using \eqref{vevinner} and the expression \eqref{Bas} for the vertex operators of the brackets, we conclude that the Yang-Mills action is given by the expectation value
\begin{equation}\label{YM SFT}
\begin{split}
S_{\rm YM}[A]&=\frac12\,\big\l\cV(A)Q\cV(A)\big\r+\frac13\,\big\l\cV^3(A)\big\r+\frac18\,\big\l\cV(A)\{\cV_2(A,A),\cV(A)\}\big\r\\
&\equiv-\frac14\int d^Dx \,F^{\mu\nu}_aF_{\mu\nu}^a\;,
\end{split}    
\end{equation}
upon observing that  $\big\l\cV(A)Q\cV_2(A,A)\big\r=0$, which can be checked by direct computation. Notably, this implies that the cubic vertex of Yang-Mills is given by $\big\l\cV^3(A)\big\r$ and the action up to cubic order is the Chern-Simons functional for $\cV(A)$. 

We conclude with some remarks on background covariance of this approach. The action \eqref{YM SFT} relies explicitly on a perturbative expansion in powers of $A_\mu^a$, as it is common in the $L_\infty$ approach to field theories and in string field theory. As such, the above action does not look geometric and gauge invariance is not manifest. Background independence is at the core of the geometric formulation of gravity, but its role in gauge theory is less apparent. To appreciate this point, we recall that the $L_\infty$ algebra $\cX^{\rm YM}$ is a vector space, implying that the field $A_\mu^a\in X^{\rm YM}_0$ is also an element of a vector space. On the other hand, gauge fields are connections, which do \emph{not} form a vector space. To reconcile these two viewpoints one should really think of $A_\mu^a\in X^{\rm YM}_0$ as a fluctuation around the trivial connection $\bar A_\mu=0$.

Although the vertex operators and action do not look geometric, they descend from the deformed BRST operator $Q_A$, which is background independent in the sense that it depends only on $\cD_\mu$. One can define vertex operators and $L_\infty$ brackets around any background gauge connection $\bar A_\mu$, as long as it is a solution of the Yang-Mills equations. Around such a background the differential is $Q_{\bar A}$, while the linear vertex operator is given by
\begin{equation}
\begin{split}
\cV_{\bar A}(A)&=\cV(A)+\cV_2(\bar A,A)\;,
\end{split}    
\end{equation}
and $\cV_2(A,A)$ is unchanged. The entire construction of this section, in particular the dictionary \eqref{Vbrackets dictionary}, still holds upon replacing $Q\rightarrow Q_{\bar A}$ and $\cV(A)\rightarrow\cV_{\bar A}(A)$. This vertex operator formalism is thus background covariant, in the sense that one has to choose a background to define $Q_{\bar A}$ and $\cV_{\bar A}(A)$, but the structure of the theory is the same for any background solution $\bar A_\mu$.

\section{Conclusions}\label{sec:conclusions}

In this paper we have constructed off-shell vertex operators for Yang-Mills theory, using the bosonic spinning particle. Upon introducing vertex operators for all elements of the $L_\infty$ complex, we have shown how the entire $L_\infty$ algebra of Yang-Mills is encoded in their commutation relations. In particular, the three-bracket $B_3(A,A,A)$ is derived from at most bilinear operators, namely $\cV(A)$ and $\cV_2(A,A)$. This suggests that vertex operators can simplify the algebraic structure of Yang-Mills theory, in that higher brackets are derived from more fundamental objects. This opens up two natural directions for the future:
\begin{itemize}
\item The $L_\infty$ algebra of Yang-Mills captures the familiar properties of gauge theories, such as their gauge algebra and consistent interactions. Upon color stripping \cite{Zeitlin:2008cc,Bonezzi:2022yuh,Bonezzi:2023lkx}, the purely kinematic space of Yang-Mills carries a vast and hidden algebraic structure \cite{Reiterer:2019dys,Bonezzi:2022bse}. This appears to be the off-shell incarnation of the color-kinematics duality needed for double copy \cite{Bern:2019prr}. Given the potential simplifications brought by vertex operators \cite{Mafra:2022wml,Fu:2018hpu,Ben-Shahar:2021doh,Cederwall:2023wxc}, it is rather compelling to investigate whether they can be used to derive this kinematic algebra in a constructive manner. We will start addressing this problem in a forthcoming paper.
\item In our analysis we have used the Hamiltonian BRST formalism in canonical quantization. Similar to string field theory, this approach makes it easier to connect the first-quantized system to the corresponding field theory, which we have exploited. It would now be beneficial to extend this connection to the Lagrangian path integral, as it is there that the worldline formalism is most advantageous. For instance, in \cite{Dai:2008bh} it was shown that the contribution of the quartic vertex to the four-gluon amplitude can be entirely captured by integrating linear vertex operators on the worldline, not needing the bilinear vertex $\cV_2(A,A)$. It would thus be particularly helpful to establish a rigorous dictionary between the operator form used in this paper and the path integral on the line. Recent results \cite{Ahmadiniaz:2021ayd,Ahmadiniaz:2021fey} using the Bern-Kosower rules suggest that this could have important applications for studying color-kinematics duality and the kinematic algebra.
\end{itemize}

\subsection*{Acknowledgements}

I would like to thank Giuseppe Casale, Christoph Chiaffrino, Felipe D\'iaz Jaramillo, Olaf Hohm and Jan Plefka for discussions and collaborations on related topics.
The work of the author is funded by the Deutsche Forschungsgemeinschaft (DFG, German Research Foundation)–Projektnummer 524744955.

\bibliographystyle{utphys}

\end{document}